\documentclass[superscriptaddress,altaffilletter,endfloats*,preprint,aps,pra,showkeys]{revtex4-1}
\usepackage{graphicx,amsmath}

\newcommand{\real}{\mathbf{Re}}

\begin{document}

\title{Terahertz Nonlinearity in Graphene Plasmons}

\author{Mohammad M. Jadidi}
\email{mmjadidi@umd.edu}
\affiliation{Institute for Research in Electronics \& Applied Physics, University of Maryland, College Park, MD 20742, USA}

\author{Jacob C. K\"onig-Otto}
\email{j.koenig-otto@hzdr.de}
\affiliation{Helmholtz-Zentrum Dresden-Rossendorf, PO Box 510119, D-01314 Dresden, Germany}
\affiliation{Technische Universit\"at Dresden, 01069 Dresden, Germany}

\author{Stephan Winnerl}
\email{s.winnerl@hzdr.de}
\affiliation{Helmholtz-Zentrum Dresden-Rossendorf, PO Box 510119, D-01314 Dresden, Germany}

\author{Andrei B. Sushkov}
\email{sushkov@umd.edu}
\affiliation{Center for Nanophysics and Advanced Materials, University of Maryland, College Park, Maryland 20742, USA}

\author{H. Dennis Drew}
\email{hdrew@umd.edu}
\affiliation{Center for Nanophysics and Advanced Materials, University of Maryland, College Park, Maryland 20742, USA}

\author{Thomas E. Murphy}
\email{tem@umd.edu}
\affiliation{Institute for Research in Electronics \& Applied Physics, University of Maryland, College Park, MD 20742, USA}

\author{Martin Mittendorff}
\email{martin@mittendorff.email}
\affiliation{Institute for Research in Electronics \& Applied Physics, University of Maryland, College Park, MD 20742, USA}

\begin{abstract}  
Sub-wavelength graphene structures support localized plasmonic resonances in the terahertz and mid-infrared spectral regimes\cite{Ju2011,Brar2013,Yan2013}. The strong field confinement at the resonant frequency is predicted to significantly enhance the light-graphene interaction\cite{Mikhailov2008,Koppens2011}, which could enable nonlinear optics at low intensity \cite{Gullans2013,Manjavacas2011,Jablan2015} in atomically thin, sub-wavelength devices \cite{Kauranen2012}.  To date, the nonlinear response of graphene plasmons and their energy loss dynamics have not been experimentally studied.  We measure and theoretically model the terahertz nonlinear response and energy relaxation dynamics of plasmons in graphene nanoribbons.  We employ a THz pump-THz probe technique at the plasmon frequency and observe a strong saturation of plasmon absorption followed by a 10 ps relaxation time. The observed nonlinearity is enhanced by two orders of magnitude compared to unpatterned graphene with no plasmon resonance.  We further present a thermal model for the nonlinear plasmonic absorption that supports the experimental results.
\end{abstract}%
\maketitle

Graphene exhibits a broadband intrinsic nonlinear optical response \cite{Mikhailov2008,Hendry2010} that has been used in mode-locking \cite{Zhang2009} and harmonic generation \cite{Hong2013}. In the optical and near-infrared regime, the nonlinear response of graphene is primarily attributed to transient Pauli blocking, which leads to an ultrafast saturable absorption and nonlinear refraction\cite{Zhang2012}.  In the terahertz regime\cite{Jnawali2013,Shi2014}, however, the nonlinear response is primarily caused by fast thermal heating and cooling of the electron population, which effects the intraband absorption\cite{Kadi2014,Mics2015}.

In the terahertz and mid-IR regime, the light-graphene interaction can be greatly increased by exploiting plasmon resonances, where the field is strongly localized and resonantly enhanced in a sub-wavelength graphene region\cite{Koppens2011}.  A dramatic enhancement of the linear absorption has been experimentally observed in isolated subwavelength graphene elements\cite{Ju2011,Brar2013,Yan2013}, and graphene-filled metallic apertures \cite{Jadidi2015} at resonant frequencies that can be controlled through the graphene dimensions and carrier concentration. Significant enhancement in the nonlinear response of graphene can be expected and has been theoretically predicted\cite{Koppens2011,Manjavacas2011,Gullans2013,Yao2014,Jablan2015}.  To date, there have been no experimental demonstrations to study this effect, or to explore the energy loss dynamics of these collective plasmonic excitations.

In this letter, we measure the nonlinear response of plasmon resonances in an array of graphene nanoribbons using THz pump-THz probe measurements with a free-electron laser that is tuned to the plasmon resonance (9.4 THz.)  We observe a resonantly-enhanced pump-induced nonlinearity in the transmission that is orders of magnitude stronger than that of unpatterned graphene.  The pump-probe measurements reveal an energy relaxation time of approximately 10 ps (measured at 20K).  We present a thermal model of the nonlinear plasmonic response that includes scattering through LA phonons and disorder-assisted supercooling, which matches both the observed timescale and power-scaling of the nonlinear response.  With the model we show that the strong pump-induced transmission change is caused by an unexpected red-shift and broadening of the resonance. Furthermore the model predicts that even greater resonant enhancement of the nonlinear response can be expected in high-mobility graphene, suggesting that nonlinear graphene plasmonic devices could be promising candidates for classical and quantum nonlinear optical processing.

\begin{figure}
  \centering
  \includegraphics[scale=1.75]{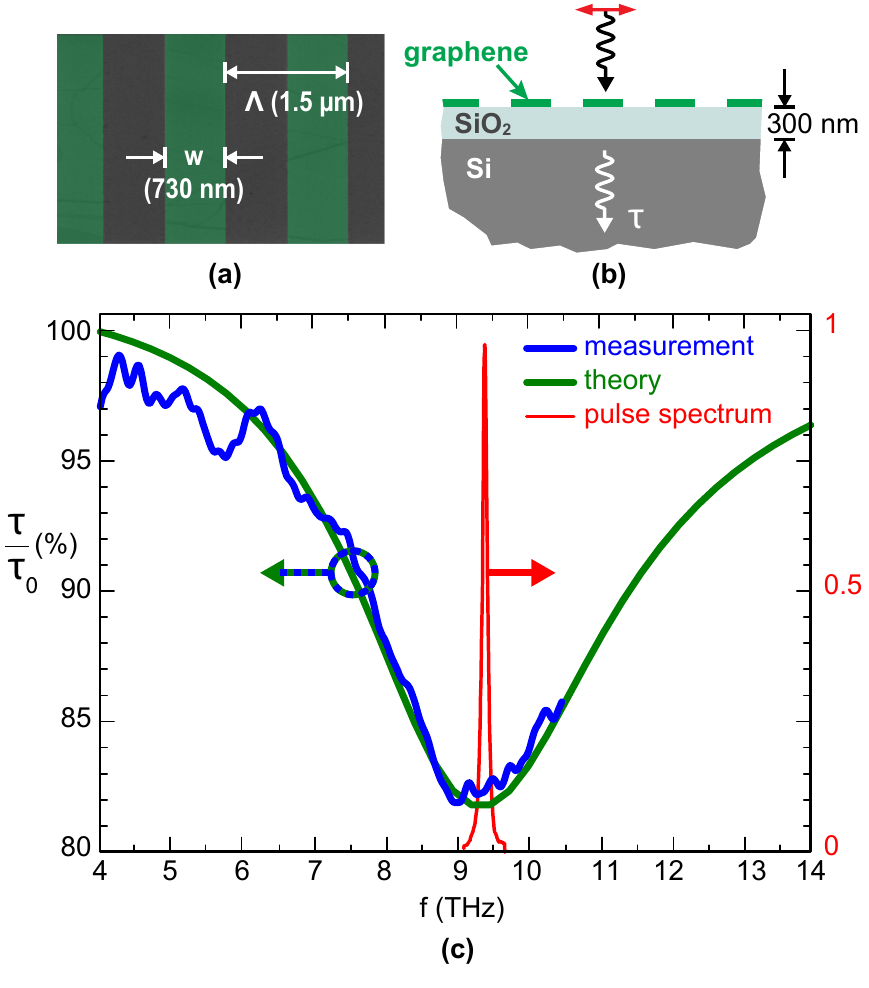}
  \caption{\textbf{Structure and transmission spectrum of graphene nano ribbons.}  (a) False color scanning electron micrograph of fabricated graphene ribbons.  (b) Cross sectional diagram of device.  (c) Measured (blue) and best fit (green) linear transmission spectrum of device, showing a decreased transmission at the plasmon frequency of 9.4 THz.  The superposed red curve shows the measured spectrum of the free electron laser pulse source that was used to observe the nonlinear response.}
  \label{fig:1}
\end{figure}

Fig.~\ref{fig:1}a-b shows the structure, dimensions, and scanning-electron micrograph of the graphene plasmonic resonant structure considered here, and Fig.~\ref{fig:1}c shows the measured room-temperature linear transmission spectrum of the sample, which exhibits a strong dip in transmission centered at 9.4 THz that is associated with plasmonic absorption of the nanoribbons.

The plasmon resonance can be approximated by assuming an equivalent sheet conductivity of the graphene ribbon array \cite{Chen2013,Cai2015} (Supplementary Equation S9),
\begin{equation}\label{eq:1}
  \sigma(\omega) =
  \frac{w}{\Lambda}\frac{D}{\pi[\Gamma-i(\omega^2-\omega_p^2)/\omega]}
\end{equation}
where $\Gamma$ is the scattering rate and $D \simeq \sqrt{\pi n}e^2v_F/\hbar$ is the Drude weight of graphene with a carrier concentration of $n$ and Fermi velocity $v_F$.  The plasmon resonant frequency is related to the Drude weight by $\omega_p^2 \equiv D w/\left[2 \Lambda^2 \epsilon_0\bar\epsilon \ln\left(\sec\left(\pi w/2\Lambda\right)\right)\right]$, where $\bar\epsilon = (\epsilon_1 + \epsilon_2)/2$ is the average of the substrate and incident dielectric constants \cite{Chen2013}.

The relative power transmission through such a conductive sheet is given by $\tau(\omega)/\tau_0 = \left|1+\sigma(\omega)/(Y_1+Y_2)\right|^{-2}$, where $\tau_0$ denotes the transmission with the graphene film absent, and $Y_j \equiv (\epsilon_0\epsilon_j/\mu_0)^{1/2}$ is the admittance of the incident ($j=1$) or substrate ($j=2$) region (see Supplementary Equation S4.)  The green curve in Fig.~\ref{fig:1}b shows the best-fit transmission spectrum calculated using this model, from which we determined the carrier concentration and graphene scattering rate to be $n = 9\times 10^{12}$ cm$^{-2}$ and $\Gamma = 23$ rad/ps, respectively at room temperature, which corresponds to a Fermi energy of 0.35 eV and carrier mobility of 1,250 cm$^2$V$^{-1}$s$^{-1}$.

\begin{figure}
  \centering
  \includegraphics[scale=1.75]{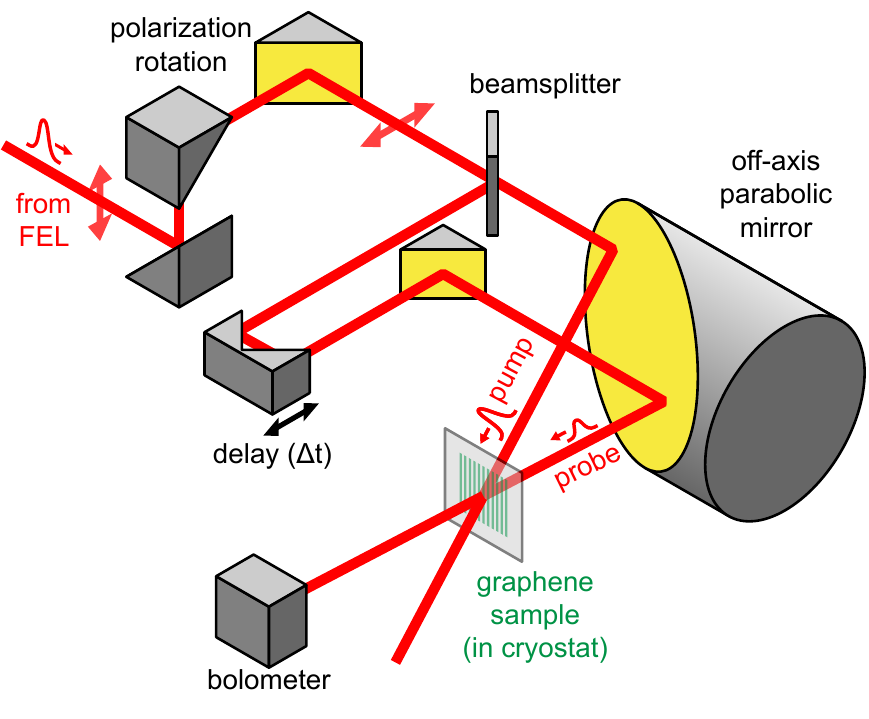}
  \caption{\textbf{Terahertz pump-probe measurement system.}  The free-electron laser (FEL) was tuned to generate 5.5 ps pulses at a carrier frequency of 9.4 THz and repetition rate of 13 MHz.  An optional reflective polarization rotation system orients the polarization perpendicular to the graphene ribbons.  The pulses were separated into parallel, co-polarized pump and probe pulses that were focused onto the graphene sample inside of a cryostat.  The transmitted probe power was measured as a function of the relative pump-probe delay $\Delta t$, which was controlled through a mechanical delay stage.}
  \label{fig:2}
\end{figure}

\begin{figure}
  \centering
  \includegraphics[scale=0.95]{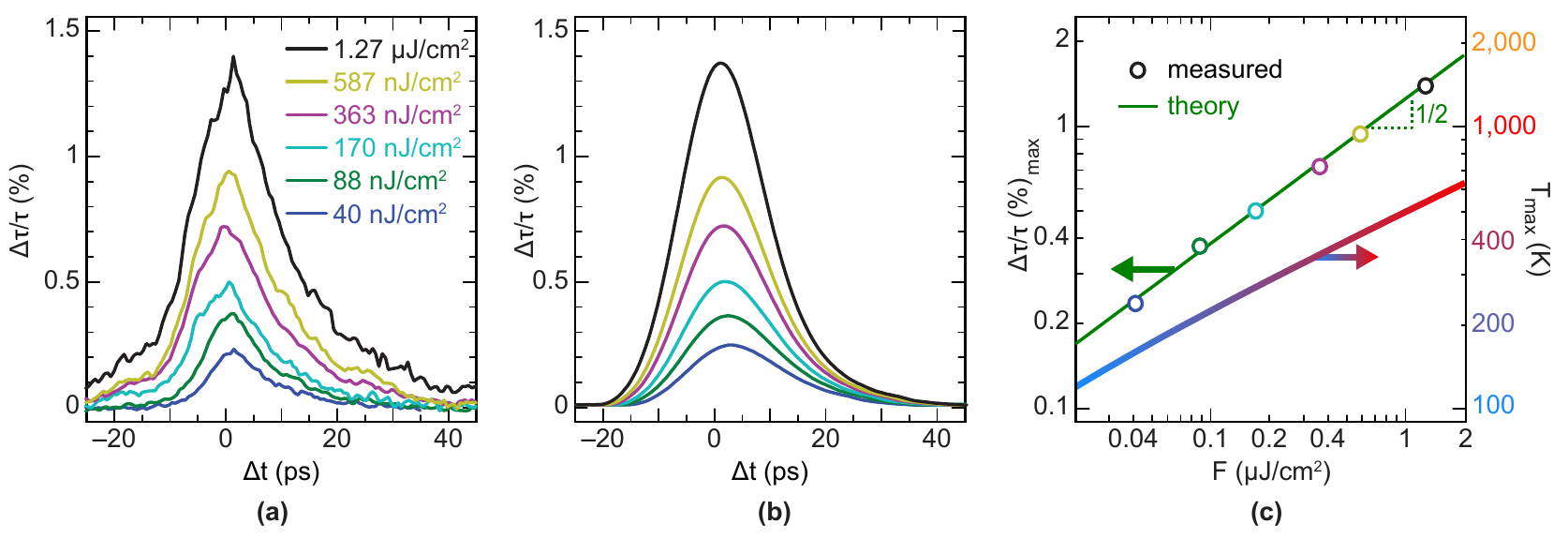}
  \caption{\textbf{Pump-probe measurements and model.} (a) Measured relative change in transmission of the probe signal for different pump fluences as a function of pump-probe time delay $\Delta t$. The positive signal indicates a decrease in absorption that becomes stronger at higher pump fluences. (b) Calculated relative change in transmission based on a nonlinear thermal model for plasmonic absorption in graphene nanoribbons that includes supercollision cooling and LA phonon scattering. (c) Measured and simulated peak of relative transmission change (left) and peak electron temperature (right) as a function of pump fluence $F$.}
  \label{fig:3}
\end{figure}

The pump-induced transmission change $\Delta \tau/\tau$ at the center of the plasmonic resonance was measured with spectrally narrow radiation (cf. Fig.~\ref{fig:1}c) in a setup that is depicted in Fig.~\ref{fig:2}. This signal, recorded as a function of the pump-probe delay $\Delta t$, is depicted in Fig.~\ref{fig:3}a for several pump fluences. In all cases, the pump causes a transient increase in transmission that is accompanied by a decrease in absorption.  The observed nonlinear response decays in the wake of the pump pulse with a time constant of $\sim$ 10 ps, which is close to the previously reported hot electron-phonon relaxation time in graphene at the measurement temperature (20 K) \cite{Winnerl2011}.

The electron temperature $T$ in the graphene evolves in response to the terahertz pump pulse with intensity $I(t)$ at the center frequency $\omega_0$ according to \cite{Song2012}
\begin{equation}\label{eq:2}
  \alpha T \frac{dT}{dt} + \beta (T^3-T_L^3) = A(\omega_0;T) I(t)
\end{equation}
where $\alpha=2\pi k_B^2 \varepsilon_F/(3\hbar^2 v_F^2)$ is the specific heat of graphene, $\beta=\zeta(3) V_D^2 \varepsilon_F k_B^3/(\pi^2 \rho \hbar^4 v_F^3 s^2 l)$ is the cooling coefficient, $T_L$ is the lattice temperature, $A(\omega_0;T)$ is the fractional absorption in the graphene, which itself depends on temperature.  $k_B$ is the Boltzmann constant,  $\rho$ is the areal mass density, $s$ is the speed of sound in graphene, $\zeta$ is the Riemann zeta function, $l$ is the electron-disorder mean free path, and $V_D$ is the acoustic deformation potential. We assume that the temperature relaxation is dominated by disorder-assisted supercollision cooling $\propto T^3$ \cite{Viljas2010,Song2012}, rather than momentum-conserving cooling\cite{Kar2014}.

The fractional absorption appearing in \eqref{eq:2} can be derived from the equivalent conductivity \eqref{eq:1} (Supplementary Equation S3),
\begin{equation}\label{eq:3}
  A(\omega_0;T) = \frac{4Y_1\real\left\{\sigma(\omega_0)\right\}}{|Y_1+Y_2+\sigma(\omega_0)|^2}
\end{equation}
where $\omega_0$ denotes the carrier frequency of the quasi-CW pump and probe pulses.

The basis of the thermal model is that the Drude weight $D$, scattering rate $\Gamma$, and plasmon frequency appearing in \eqref{eq:1} implicitly depend upon the electron temperature $T$, which increases when the incident pump pulse is absorbed in the graphene layer.

The temperature-dependent Drude weight \cite{DasSarma2011,Frenzel2014} and plasmon frequency (supplementary Section S3) are calculated as 
\begin{align}
  \label{eq:5}
  D(T) &= \frac{2e^2}{\hbar^2}k_BT\ln\left[2\cosh\left(\frac{\mu(T)}{k_BT}\right)\right] \\
  \label{eq:6}
  \omega_p^2(T) &= \frac{D(T)w}{2 \epsilon_0\bar\epsilon \Lambda^2 \ln\left(\sec\left(\pi w/2\Lambda\right)\right)}
\end{align}

The scattering rate $\Gamma$ also varies with temperature, both because of temperature-dependent scattering from long-range Coulomb impurities and longitudinal acoustic (LA) phonons \cite{Chen2008}.
\begin{equation}\label{eq:7}
  \Gamma(T) = \frac{\Gamma_0 \varepsilon_F}{\mu(T)} + \frac{k_B T \varepsilon_F V_D^2}{4\hbar^3v_F^2 \rho s^2} \\
\end{equation}
The second term in \eqref{eq:7} describes the temperature dependent LA phonon scattering, which was essential in order to match the observed fluence dependence of the nonlinear response, shown in Fig.~\ref{fig:3}c (see Supplementary Section S4.)

The results from the thermal model (Fig.~\ref{fig:3}b) are in close agreement with the experimental data (Fig.~\ref{fig:2}a), and correctly predict the 10 ps response time. The increased transmission is a result of a decreased plasmon frequency, which is caused by a reduced value of the chemical potential at elevated electron temperatures (cf. Eq. \eqref{eq:4} and \eqref{eq:5}), and a broadening of the resonance caused by a faster scattering rate. Fig.~\ref{fig:3}c plots the peak value of the (observed and calculated) transient response as a function of the incident pump fluence $F$, which shows an approximate $F^{1/2}$ dependence.  Along the right axis, we plot the corresponding simulated peak electron temperature as a function of fluence, showing the expected $F^{1/3}$ dependence.  The observed power scaling was best matched by assuming supercollision cooling as the single dominant cooling mechanism, together with temperature-dependent momentum scattering through LA phonons\cite{DasSarma2011,Chen2008}.

Only two free parameters were used in the numerical simulations: the acoustic deformation potential $V_D$ and the electron disorder mean free path $l$, which together control the strength of dominant cooling and scattering mechanisms.  The observed $F^{1/2}$ power scaling seen in Fig.~\ref{fig:3}c was matched by choosing $V_D = 11$ eV, which is consistent with values reported in the literature for similar graphene \cite{Kar2014,McKitterick2015}.  The mean free path $l$ was adjusted to match the overall magnitude of the nonlinearity, from which we obtained $l=2$ nm -- which is smaller than that expected from the scattering rate, but consistent with other recent experimental measurements of cooling in large-area graphene\cite{McKitterick2015}.  The origin of this discrepancy remains to be explained.

\begin{figure}
  \centering
  \includegraphics[scale=1.75]{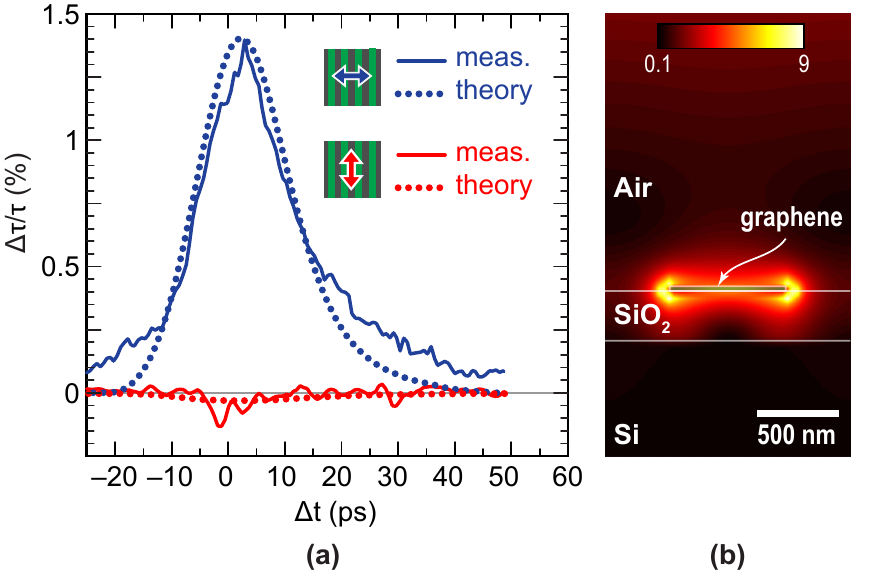}
  \caption{\textbf{Plasmonic enhancement of nonlinearity.}  (a) Comparison of normalized change in transmission for two different polarizations.  The blue curves show the measured and simulated pump-probe response when the pump and probe were polarized perpendicular to the graphene nanoribbons, thereby exciting the plasmon.  The red curves show the measured and simulated response for the same incident pump fluence, but opposite polarization, where there is no plasmonic excitation, and the nonlinear response is correspondingly much lower.  (b) Electric field profile at the resonant frequency, calculated using a (linear) finite element time domain method with a normally incident wave from above, showing the field-enhancement at the graphene surface.  The color indicates the electric field intensity $|\mathbf{E}|^2$, relative to that of the incident plane wave, showing a nearly 9-fold intensity enhancement at the graphene surface. }
  \label{fig:4}
\end{figure}

To confirm the plasmonic enhancement of the nonlinearity, we repeated the pump-probe measurements with the pump and probe co-polarized in the direction parallel to the graphene ribbons, thus ensuring that the plasmons are not excited.  Fig.~\ref{fig:4}a compares the measurements from the two polarization cases for the same incident pump fluence and frequency.  The measured nonlinearity is far stronger when the plasmons are excited than for the opposite polarization, consistent with the thermal predictions.  Fig.~\ref{fig:4}b shows the electric field profile at the plasmon resonance calcuated using (linear) finite element simulations, showing the dramatic field enhancement that occurs near the graphene sheet, which is responsible for the enhanced nonlinearity.

\begin{figure}
  \centering
  \includegraphics[scale=1.75]{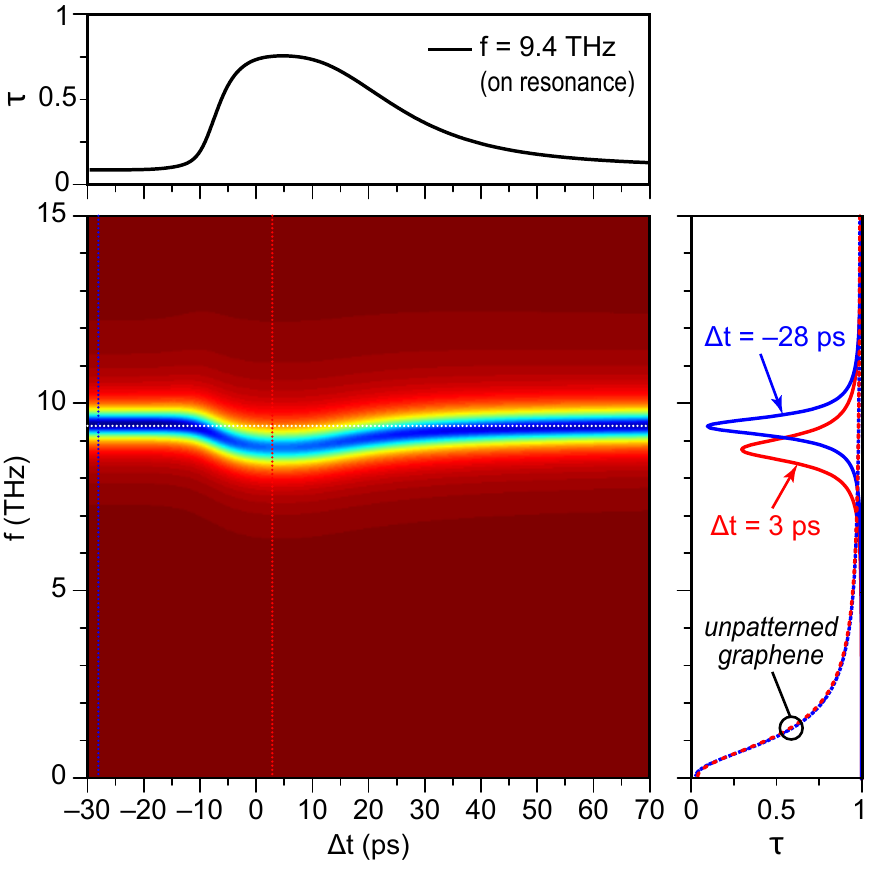}
  \caption{\textbf{Transient change in plasmon transmission spectrum.}  Numerically predicted change in transmission as a function of frequency $f$ and time $\Delta t$, calculated assuming a higher graphene mobility of 25,100 cm$^2$V$^{-1}$s$^{-1}$.  The pump pulse causes a transient red-shift and broadening of the plasmon resonance, as shown by the two vertical sections plotted on the right.  The dashed curves indicated in the right panel show the calculated Drude response for an unpatterned graphene sheet, which shows no plasmon resonance, and very little pump-induced change in transmission.  A signal tuned to the resonant frequency would experience a corresponding transient increase in transmission, as shown in the horizontal section plotted on the top.}
  \label{fig:5}
\end{figure}

Fig.~\ref{fig:5} presents a calculation of how this nonlinearity would be further enhanced by employing higher quality graphene nanoribbons with a mobility of 25,100 cm$^2$V$^{-1}$s$^{-1}$\cite{Wang2013}.  The calculated power transmission is shown as a function of frequency (in the vertical direction) and time (in the horizontal direction), assuming an input fluence of 1.27 $\mu$J/cm$^2$.  The nonlinear transmission is caused by a transient red-shift of the plasmon frequency and broadening of the plasmon linewidth, causing a pump-induced change in transmission of order unity.

To conclude, the temperature dependent absorption, cooling, and scattering of hot electrons in graphene causes a nonlinear response to terahertz waves.  Using terahertz pump-probe measurements, we show that when graphene is patterned into sub-wavelength structures that exhibit a plasmon resonance, this nonlinearity is greatly enhanced at the resonant frequency.  This enhanced nonlinearity is caused by a stronger on-resonance absorption, followed by a spectral red-shift and broadening of the plasmon resonance with electron temperature.  We provide a thermal model that explains the observed nonlinear enhancement, and sheds light on the dominant cooling and scattering mechanisms for hot electrons collectively excited in a graphene plasmon.  The theory predicts that in higher-mobility graphene the nonlinearity in transmission could approach unity, enabling high-speed terahertz-induced switching or modulation.

\bibliographystyle{naturemag}
\bibliography{Jadidi-NPhys-2015}

\section*{Methods}

\textbf{Device Fabrication:}  The plasmonic devices were fabricated using CVD-grown monolayer graphene that was transferred onto a oxidized silicon substrate and diced into an 8 mm square chip.  The substrate resistivity was 250 $\Omega\cdot$cm, and the oxide was 300 nm thick.  Graphene ribbons with width $w=$730 nm and period $\Lambda=$1.5 $\mu$m were patterned using electron-beam lithography with a PMMA resist and oxygen plasma etch to remove the graphene from the exposed areas.  The graphene grating covered a region of 1.5 $\times$ 1.5 mm.

\textbf{FEL pump-probe setup:} The free electron laser was tuned to produce 5.5 ps pulses at a center frequency of 9.4 THz and a repetition rate of 13 MHz.  The beam was split into a pump and probe beams that were delayed relative to one another using a mechanical delay line. In all measurements the pump and probe beams were co-polarized, but the state of polarization could be set to be either perpendicular or parallel to the graphene ribbons, in order to control whether or not the plasmon was excited, respectively.  The pump and probe beams were overlapped and focused using an off-axis parabolic mirror onto the graphene ribbon array located at the focus.  The sample was cooled to a (lattice) temperature of 20 K for all of the pump-probe measurements.  The emerging pump beam was extinguished while the transmitted probe beam was measured using a cryogenically cooled bolometer as a function of the pump-probe delay $\Delta t$.

\textbf{Nonlinear Model:}  Because of the fast electron-electron scattering time, we assume that the electron population maintains a Fermi distribution, with a temperature that evolves in response to the terahertz pump pulse according to \eqref{eq:2}.  The total electron population $n$ must remain constant as the electrons heat and cool, which defines the following implicit relationship between the electron temperature and the chemical potential $\mu(T)$,
\begin{equation}\label{eq:4}
  n = \int\limits_{-\infty}^{\infty} \frac{\nu(E)dE}{1+\exp\left[\dfrac{E-\mu(T)}{k_BT}\right]}
\end{equation}
where $\nu(E)$ is the density of states in graphene.  For a given temperature $T$, \eqref{eq:4} can be numerically solved to determine the associated chemical potential $\mu(T)$.

The electron temperature transient $T(t)$ was found by numerically integrating \eqref{eq:2} using Euler's method with a time-step of 0.5 ps, assuming a 5.5 ps Gaussian input pulse $I(t)$.  At each time-step, the chemical potential, Drude weight, plasmon frequency, conductivity, and fractional absorption for the subsequent timestep were adjusted based upon \eqref{eq:4}, \eqref{eq:5}, \eqref{eq:6}, \eqref{eq:1} and \eqref{eq:3}, respectively.

Knowing the instantaneous temperature transient $T(t)$, the fractional change in probe transmission is then numerically computed as a function of the pump-probe delay $\Delta t$ using a correlation integral.

\textbf{Finite Element Calculation:}  The field-enhancement in the graphene was simulated using frequency-domain finite element calculations, performed on a unit cell with periodic boundary conditions, assuming normally incident plane wave excitation. The graphene was modeled as a thin (40 nm) anisotropic Drude conductor, with parameters adjusted to match the observed room-temperature transmission spectrum.

\section*{Author Contributions}
M. M. Jadidi, S. Winnerl,  H. D. Drew, T. E. Murphy, and M. Mittendorff conceived the experiments. M. M. Jadidi fabricated the device and conceived the theory for the thermal model. M. M. Jadidi and A. B. Sushkov carried out the FTIR measurements. J. Otto and S. Winnerl conducted the FEL pump-probe measurements.  All authors contributed to the manuscript.

\section*{Acknowledgements}
This work was sponsored by the US ONR (N000141310865) and the US NSF (ECCS 1309750). Support from P. Michel and the FELBE-team is gratefully acknowledged.

\end{document}